\catcode`\@=11  % This allows us to modify PLAIN macros.
\def\answ{o}    %  specify  o,t, p, or n
\def\onecol{o }      %      o = one column compact preprint format
\def\twocol{t }      %      t = two column PRD format
\def\nofig{n }       %      n = one column compact but no figure inset
\def\prepri{p }      %      p = one column preprint style
\documentstyle[aps,eqsecnum,floats,epsf]{revtex}

\begin{document}
\draft
%---------------------------------------------------------------------
% for two column style turn on
\if\answ\twocol
      \twocolumn[\hsize\textwidth\columnwidth\hsize\csname
             @twocolumnfalse\endcsname
\fi
%---------------------------------------------------------------------

\title{Constraint propagation in the family of ADM systems}
\author{Gen Yoneda ${}^\dagger$ and Hisa-aki Shinkai ${}^\ddagger$}
\address{
{\tt yoneda@mse.waseda.ac.jp} ~~~ {\tt shinkai@gravity.phys.psu.edu}
\\
${}^\dagger$ Department of Mathematical Sciences, Waseda University,
Shinjuku, Tokyo,  169-8555, Japan
\\
${}^\ddagger$ Centre for Gravitational Physics and Geometry,
104 Davey Lab., Department of Physics,\\
The Pennsylvania State University,
University Park, Pennsylvania 16802-6300, USA
}
%\date{\today}
\date{March 8, 2001, gr-qc/0103032, CGPG-01/3-2}
\maketitle

%---------------------------------------------------------------------
\begin{abstract}
%---------------------------------------------------------------------
% for two column style turn on
\if\answ\twocol
     \widetext
\fi
%---------------------------------------------------------------------
%=====================================================================
%<<<<<<<<<<<<< ABSTRACT >>>>>>>>>>>>>>>%
%=====================================================================
The current important issue in numerical relativity is to determine
which formulation of the Einstein equations provides us with stable and
accurate simulations.
Based on our previous work on ``asymptotically constrained" systems,
we here present constraint propagation equations
and their eigenvalues for the Arnowitt-Deser-Misner
(ADM) evolution equations with additional
constraint terms (adjusted terms) on the right hand side.
We conjecture that the system is robust against violation
of constraints if
the amplification factors (eigenvalues of Fourier-component
of the constraint propagation equations)
are negative or pure-imaginary.
We show such a system can be obtained by choosing multipliers of
adjusted terms.
Our discussion covers Detweiler's proposal (1987)
and Frittelli's analysis (1997),
and we also mention the so-called conformal-traceless ADM systems.
\end{abstract}
%\pacno{04.20.Cv}
%\multicols{2}

%---------------------------------------------------------------------
% for two column style turn on
\if\answ\twocol
     \vskip 2pc]
     \narrowtext
\fi
%---------------------------------------------------------------------

%=====================================================================
\section{Introduction}
%=====================================================================
The effort to solve the Einstein equations numerically
-- so-called Numerical Relativity -- is now providing an
interesting bridge
between mathematical relativists and numerical relativists.
Most of the simulations have been performed using
the Arnowitt-Deser-Misner
(ADM) formulation\cite{adm} or a modified version.
However,
the ADM formulation has not been proven to be a well-posed system,
since its evolution
equations do not present a hyperbolic form in its original/standard
formulation.

Most simulations are performed using the ``free evolution"
procedures:
(1) solve the Hamiltonian and momentum constraints to prepare the
initial data,
(2) integrate the evolution equations by fixing gauge conditions,
and (3) monitor the accuracy/stability by evaluating the constraints.
Many trials have been made in the last few decades, but we have not
yet obtained a  perfect recipe
for long-term stable evolution of the Einstein equations.
Here we consider the
problem through the form of the equations.

One direction in the community is to rewrite the Einstein
evolution equations into
a hyperbolic form and to apply it to numerical
simulations \cite{hypref}.
This is motivated by the fact that
we can prove well-posedness for the evolution of several systems
if they have a certain kind of hyperbolic feature.
The authors
recently derived \cite{ysPRL,ysIJMPD} three levels of
hyperbolic system of the Einstein equations using Ashtekar's
connection variables\cite{Ashtekar}
\footnote{We derived weakly, strongly ($=$ diagonalizable)
and symmetric
hyperbolic systems.
The mathematical inclusion relation is
\begin{center}
weakly hyperbolic $\ni$ strongly hyperbolic $\ni$
symmetric hyperbolic.\\
\end{center}
See details in \cite{ysIJMPD}.
%Note that the ADM formulation is not in the category of
%weakly hyperbolic system.
}, and compared them numerically \cite{ronbun1}.
We found that (a) the three levels of hyperbolicity
can be obtained by adding
constraint terms and/or imposing gauge conditions,
(b) there is no drastic difference in
the accuracy of numerical evolutions in these three, and
(c) the symmetric hyperbolic system
is not always the best for reducing numerical errors.
Similar results regarding to (a) and (b) are
reported by Hern \cite{HernPHD} based on the Frittelli-Reula
formulation \cite{FR96}.

What are, then, the criteria for predicting the stable
evolutions of a system?
Inspired by the  ``$\lambda$-system"  proposal \cite{BFHR},
we have considered a so-called
   ``asymptotically constrained" system,   that
is, a system robust against the violation
of the constraints \cite{SY-asympAsh}.
The fundamental idea of the ``$\lambda$-system"
is to introduce artificial flow onto the constraint surface.
However, we also found that
such a feature can be obtained simply by adding constraint terms
to the evolution equations which we named ``adjusted systems"
\cite{ronbun2}.
We explained the reason why this works by analyzing the
evolution equations of the constraints
(the propagation of
the constraints).  We proposed that the stablity of the system
can be predicted by analyzing the eigenvalues (amplification factors)
of the constraint
propagation equations (We describe this in detail in \S \ref{sec2}).
%We systematically treated the process of adding constraint terms
%(we named them ``adjusted systems"),
We confirmed
that our proposal works both in Maxwell and Ashtekar
systems \cite{ronbun2}.

The purpose of this article is to apply our proposal
to the ADM system(s).
Especially, we consider the ``adjusting process"
(adding constraints in RHS of evolution equations),
and the resultant changes to the eigenvalues of the
constraint propagation systems.
This adjusting process can be seen in many constructions of hyperbolic
systems in the references.
In fact, the {\it standard ADM} for numerical relativists
is the version which was introduced by York \cite{ADM-York},
where the {\it original ADM} system \cite{adm}
  has already been adjusted using the Hamiltonian constraint
(see more detail in \S \ref{secADM}).
The advantage of the {\it standard ADM} system is reported by
{}Frittelli\cite{Fri-con} from
the point of the hyperbolicity and the characteristic
propagation speed of the constraints.
Our discussion extends her analysis to the amplification factors.

One early effort of the adjusting mechanism was presented by
Detweiler\cite{detweiler}.
Our study also includes his system, and shows that
this system actually works as desired for a certain
choice of parameter (\S \ref{secADJADM}).
We also study the same procedure for the
``conformal-traceless" ADM (CT-ADM) formulations \cite{SN,BS}
which is recently the most popular system in numerical simulations
(\S \ref{secCTADM}).

% we only consider the vacuum case here,
The analysis in the text is for perturbational
violation on a flat background.
{}Further applications are available,
but we will discuss them in future reports.
In the Appendix, we also give numerical
demonstrations of the adjusted-ADM
systems discussed in the text.

%=====================================================================
\section{Constraint propagation and ``adjusted system"} \label{sec2}
%=====================================================================
We begin by reviewing the background of
``adjusted systems" and our conjecture.

%\subsection{Constraint propagation}
The notion of the evolution equations of the constraints is often
discussed from the point of whether they form a first class system or not.
{}Fortunately, the constraints in the (original/standard) ADM
formulation are known to form a first class system.
Due to this fact, numerical relativists only need to monitor
the violation of the Hamiltonian and momentum
constraints during the free evolution of the initial data.

Our essential idea here is to feed this procedure back into the
evolution equations.
That is,  we adjust the
system's evolution equations by characterizing the
constraint propagation in advance.
Let us describe the procedure in a general form.  Suppose we have
a set of dynamical variables, $u^a (x^i,t)$,
and its evolution equations,
\begin{equation}
\partial_t u^a = f(u^a, \partial_i u^a, \cdots), \label{Deq}
\end{equation}
which should satisfy a set of constraints,
$C^\rho(u^a, \partial_i u^a, \cdots) \approx 0$.
The evolution equation for $C^a$ can be written as
\begin{equation}
\partial_t C^\rho = g(C^\rho, \partial_i C^\rho, \cdots). \label{Ceq}
\end{equation}

We can perform two main types of analysis analysis on (\ref{Ceq}):
\begin{enumerate}
\item
If (\ref{Ceq}) is in a first order form
(that is, only includes first-order spatial derivatives),
then the level of hyperbolicity and the characteristic speeds
(eigenvalues $\lambda^l$ of the principal matrix)
will definitely deterine the stability of the system.
We expect mathematically rigorous well-posed features for
strongly or symmetric
hyperbolic systems, and the characteristic speeds suggest to
us satisfactory criteria
for stable evolutions if they are real, and under the propagation
speed of the  original variables, $u^a$,
and/or within the causal region of the numerical integration
scheme applied.
\item
On the other hand, the Fourier transformed-(\ref{Ceq}),
\begin{equation}
\partial_t \hat{C}^\rho = \hat{g}(\hat{C}^\rho), \label{CeqF}
\end{equation}
where $
%\lambda(x,t)^\rho
%=\displaystyle{\int} \hat{\lambda}(k,t)^\rho\exp(ik\cdot x)d^3k,
C^\rho(x,t)
={\int} \hat{C}^\rho(k,t)\exp(ik\cdot x)d^3k,$
also characterizes the
evolution of the constraints independently of
its hyperbolicity.
As we  have proposed and confirmed in \cite{ronbun2}, the
set of eigenvalues $\Lambda^i$ of the coefficient matrix in
(\ref{CeqF}) provides a kind of {\it amplification factor} of the
constraint propagation, and predicts the increase/decrease of the
violation of the constraints if it exists.
More precisely, we showed in
\cite{ronbun2} that
\begin{itemize}
\item[]
if the eigenvalues of (\ref{CeqF}) (a) have a {\it negative}
real-part, or (b) are {\it non-zero} ({\it pure-imaginary})
eigenvalues, then we see
more stable evolutions than a system which does not.
\end{itemize}
This is because the constraints are damped
if the eigenvalues
are negative, and are propagating away
if the eigenvalues
are pure imaginary.
We found heuristically that the system becomes more stable (accurate) when
as much  $\Lambda$s satisfies the above criteria and/or as
large magnitude of
$\Lambda$s away from zeros.
(Examples in \cite{ronbun2} are of the plane wave
propagation  in the Maxwell
system and the Ashtekar system.)
We remark that this eigenvalue analysis requires that we fix
a particular
background metric for the situation we consider, since the
amplification factor depends on the dynamical variables $u^a$.
\end{enumerate}

%----------------------------------------------------------
%\subsection{Adjusted system}
The above features of the constraint propagation, (\ref{Ceq}),
will change when we modify the original evolution equations.
Suppose we adjust the RHS of (\ref{Deq})
by adding the constraints,
\begin{equation}
\partial_t u^a = f(u^a, \partial_i u^a, \cdots)
+ F(C^\rho, \partial_i C^\rho, \cdots), \label{DeqADJ}
\end{equation}
then (\ref{Ceq}) will also be modified as
\begin{equation}
\partial_t C^\rho = g(C^\rho, \partial_i C^\rho, \cdots)
+ G(C^\rho, \partial_i C^\rho, \cdots). \label{CeqADJ}
\end{equation}
By taking the characteristic speed of (\ref{CeqADJ}) and the
amplification factor of the Fourier transformed-(\ref{CeqADJ}),
the predicted stability of the system (\ref{DeqADJ}) becomes
different to that of  the original system, (\ref{Ceq}).

Our proposed ``adjusted system" is obtained by finding a
certain functional form of
$F(C^\rho, \partial_i C^\rho, \cdots)$ in (\ref{DeqADJ}) so as to get a
more stable prediction in the analysis of the  eigenvalues
$\lambda^l$ and $\Lambda^i$.
In the following discussion, we show two eigenvalues
$\lambda^l$
   and $\Lambda^i$ for each ADM system.
We remark again that the term `characteristic speed' here
is not for the dynamical
equation (\ref{Deq}), but for the constraint propagation
equations (\ref{Ceq}).

%=====================================================================
\section{Standard ADM system} \label{secADM}
%=====================================================================
\subsection{Standard ADM system and its constraint propagation}
We start by analyzing the standard ADM system.
By ``standard ADM" we mean here the most widely adopted system, due to
York \cite{ADM-York},
with evolution equations
\begin{eqnarray}
\partial_t \gamma_{ij}&=&
-2\alpha K_{ij}+\nabla_i\beta_j+\nabla_j\beta_i, \label{admevo1}
\\
\partial_t K_{ij} &=&
\alpha R^{(3)}_{ij}+\alpha K K_{ij}-2\alpha K_{ik}{K^k}_j
-\nabla_i\nabla_j \alpha
+(\nabla_i \beta^k) K_{kj} +(\nabla_j \beta^k) K_{ki}
+\beta^k \nabla_k K_{ij},  \label{admevo2}
%-\alpha \Lambda \gamma_{ij}
\end{eqnarray}
and constraint equations
\begin{eqnarray}
{\cal H}&:=& R^{(3)}+K^2-K_{ij}K^{ij}, \label{admCH} %-2\Lambda
\\
{\cal M}_i&:=& \nabla_j {K^j}_i-\nabla_i K,  \label{admCM}
\end{eqnarray}
where $(\gamma_{ij}, K_{ij})$ are the induced three-metric
and the extrinsic
curvature, $(\alpha, \beta_i)$ are the lapse function
and the shift covector,
$\nabla_i$ is the covariant derivative adapted to $\gamma_{ij}$,
and $R^{(3)}_{ij}$ is the three-Ricci tensor.

The constraint propagation equations,
which are the time evolution equations
of the Hamiltonian constraint (\ref{admCH}) and
the momentum constraints (\ref{admCM}), can be written as
\begin{eqnarray}
\partial_t {\cal H}&=&
\beta^j (\partial_j {\cal H})
-2\alpha \gamma^{ij}(\partial_i {\cal M}_j)
+2\alpha K{\cal H}
+\alpha(\partial_l \gamma_{mk})
   (2\gamma^{ml}\gamma^{kj}-\gamma^{mk}\gamma^{lj}){\cal M}_j
-4\gamma^{ij} (\partial_j\alpha){\cal M}_i,
\label{d11H}
\\
\partial_t {\cal M}_i&=&
-(1/2)\alpha (\partial_i {\cal H})
+\beta^j (\partial_j {\cal M}_i)
+\alpha K {\cal M}_i
-(\partial_i\alpha){\cal H}
-\beta^k\gamma^{jl}(\partial_i\gamma_{lk}){\cal M}_j
+(\partial_i\beta_k)\gamma^{kj}{\cal M}_j.
\label{d11M}
\end{eqnarray}
The simplest derivation of (\ref{d11H}) and (\ref{d11M})
is by using the Bianchi identity, which
can be seen in Frittelli \cite{Fri-con}.
(Note that $C$ in \cite{Fri-con} is half our ${\cal H}$,
and we have corrected typos in eq.(11) in \cite{Fri-con}).

The characteristic part of (\ref{d11H}) and (\ref{d11M})
can be extracted as
\begin{equation}
\partial_t \left( \matrix{ {\cal H} \cr {\cal M}_i }\right)
\simeq
\left( \matrix{ \beta^l & -2\alpha\gamma^{il} \cr
-(1/2)\alpha\delta^l_i &  \beta^l \delta^j_i }\right)
\partial_l \left( \matrix{ {\cal H} \cr {\cal M}_j }\right)
=: P^l \, \partial_l \left( \matrix{ {\cal H} \cr {\cal M}_j }\right),
\label{admconprop}
\end{equation}
which indicates that the characteristic speeds
(eigenvalues of the characteristic matrix, $P^l$) are
\begin{eqnarray}
\lambda^l&=&
(\beta^l,\beta^l,\beta^l\pm\alpha\sqrt{\gamma^{ll}})
\quad (\mbox{no sum over } l).
\label{Ceigen1}
\end{eqnarray}
Since $\mbox{rank}(P^l-\beta^l)=2$,
the matrix $P^l$ is diagonalizable,
but not the symmetric.

Simply by inserting (1/2) in front of ${\cal H}$ above, we obtain
\begin{equation}
\partial_t \left( \matrix{ {\cal H}/2 \cr {\cal M}_i }\right)
\simeq
\left( \matrix{ \beta^l & -\alpha\gamma^{il} \cr
                 -\alpha\delta^l_i &  \beta^l \delta^j_i }\right)
   \partial_l \left( \matrix{ {\cal H}/2 \cr {\cal M}_j }\right);
\label{admconprop2}
\end{equation}
the characteristic matrix becomes symmetric
(with the same eigenvalues).
This is a feature of the standard ADM system that
was pointed out by Frittelli.
(Actually ${\cal H}/2$ is the form
originally given by the
Lagrangian formulation.)

%=====================================================================
%\subsection{Amplification factors of the constraint propagation}
\subsection{Amplification factors on the Minkowskii background}
As a first example, we consider the perturbation
of the Minkowskii spacetime:
$\alpha=1$, $\beta^i=0$, $\gamma_{ij}=\delta_{ij}$.
%$+{}^{(1)\!}h_{ij}$
By taking the linear order contribution,
(\ref{d11H}) and (\ref{d11M})
are reduced to
\begin{eqnarray}
\partial_t
\left(
\matrix{{}^{\!(1)\!\!} \hat{{\cal H}} \cr {}^{\!(1)\!\!} \hat{{\cal M}}_i}
\right)
=
\left(\matrix{0&-2ik_j\cr -(1/2) ik_i&0 }\right)
\left(
\matrix{{}^{\!(1)\!\!} \hat{{\cal H}} \cr {}^{\!(1)\!\!} \hat{{\cal M}}_j}
\right), \label{admMin}
\end{eqnarray}
in Fourier components. The eigenvalues of the coefficient
matrix of (\ref{admMin}), which we call {\it amplification factors}, become
%\begin{verbatim}
%Eigenvalues[{{0,-2*I*k1,-2*I*k2,-2*I*k3},
%        {-(1/2)*I*k1,0,0,0},
%        {-(1/2)*I*k2,0,0,0},
%        {-(1/2)*I*k3,0,0,0}}]
%\end{verbatim}
\begin{equation}
\Lambda^l=(0,0, \pm i\sqrt{k^2}), \label{Beigen_standardADM}
\end{equation}
where $k^2=k^2_x+k^2_y+k^2_z$. These factors will be compared with
others later, but we note that
  the real parts of all the $\Lambda$s are zero.

%=====================================================================
\section{Adjusted ADM systems} \label{secADJADM}
%=====================================================================
%=====================================================================
\subsection{Adjustments}
Generally, we can write the adjustment terms to
(\ref{admevo1}) and (\ref{admevo2})
using (\ref{admCH}) and (\ref{admCM}) by the following combinations
(using up to the first derivative of constraints),
\begin{eqnarray}
&\mbox{adjustment term of }\quad
\partial_t \gamma_{ij}:& \quad
+P_{ij} {\cal H}
+Q^k{}_{ij}{\cal M}_k
+p^k{}_{ij}(D_k {\cal H})
+q^{kl}{}_{ij}(D_k {\cal M}_l), \label{adjADM1}
\\
&\mbox{adjustment term of}\quad
\partial_t K_{ij}:& \quad
+R_{ij} {\cal H}
+S^k{}_{ij}{\cal M}_k
+r^k{}_{ij} (D_k{\cal H})
+s^{kl}{}_{ij}(D_k {\cal M}_l), \label{adjADM2}
\end{eqnarray}
where $P, Q, R, S$ and $p, q, r, s$ are multipliers  (please do not
confuse $R_{ij}$ with
three Ricci curvature  that we write as $R^{(3)}_{ij}$).
Since this expression is too general,
we mention some restricted cases below.

We remark that our starting system,
(\ref{admevo1}) and (\ref{admevo2}),
is the {\it standard ADM} system for numerical relativists
introduced by York \cite{ADM-York}.
This expression can be obtained from
the originally formulated canonical expression by ADM \cite{adm},
but in that
process the Hamiltonian constraint equation is used to eliiminate the
three dimensional Ricci scalar.
Therefore
the {\it standard ADM} is already adjusted from the
{\it original ADM} system.
We start our comparison with this point.

%=====================================================================
\subsection{Original ADM vs Standard ADM}\label{section_OriSta}
{}Frittelli's adjustment analysis \cite{Fri-con}  can be written in
terms of (\ref{adjADM1}) and (\ref{adjADM2}), as
\begin{eqnarray}
R_{ij}=(1/4) (\mu-1) \alpha \gamma_{ij}, \label{originalADMadjust}
\end{eqnarray}
where $\mu$ is a constant
and set other multiplier zero. Here $\mu=1$ corresponds to the
standard ADM (no adjustment, since $R_{ij}=0$),
and $\mu=0$ to the original ADM (without any
adjustment to the canonical formulation by ADM).

Keeping the multiplier (\ref{originalADMadjust}) in mind,
we here discuss the
case of non-zero $R_{ij}, S^k{}_{ij}$ (and all other multipliers zero) case.
The constraint propagation equations become
\begin{eqnarray}
\partial_t {\cal H}&=&
\beta^j (\partial_j {\cal H})
-2\alpha\gamma^{ij}(\partial_i {\cal M}_j)
+2\alpha K{\cal H}
+\alpha(\partial_l \gamma_{mk})
   (2\gamma^{ml}\gamma^{kj}-\gamma^{mk}\gamma^{lj}){\cal M}_j
-4\gamma^{ij} (\partial_j\alpha){\cal M}_i
\nonumber \\&&
+2KR {\cal H}
-2K^{ij}R_{ij} {\cal H}
+2K\gamma^{ij}S^k{}_{ij}{\cal M}_k
-2K^{ij}S^k{}_{ij}{\cal M}_k,
\label{tcha}
\\
\partial_t {\cal M}_i&=&
-(1/2)\alpha (\partial_i {\cal H})
+\beta^j (\partial_j {\cal M}_i)
+\alpha K {\cal M}_i
-(\partial_i\alpha){\cal H}
-\beta^k\gamma^{jl}(\partial_i\gamma_{lk}){\cal M}_j
+(\partial_i\beta_k)\gamma^{kj}{\cal M}_j
\nonumber \\&&
%\\&& % itK
+\gamma^{kj}(\partial_jR_{ki}) {\cal H}
-\gamma^{jk}(\partial_iR_{jk}) {\cal H}
+R^j{}_i (\partial_j{\cal H})
-R_{jk}\gamma^{jk} (\partial_i{\cal H})
\nonumber \\&&
+\gamma^{lj}(\partial_jS^k{}_{li}){\cal M}_k
-\gamma^{jl}(\partial_iS^k{}_{jl}){\cal M}_k
+S^{kj}{}_i(\partial_j{\cal M}_k)
-\gamma^{jl}S^k{}_{jl}(\partial_i{\cal M}_k)
\nonumber \\&& % tK
+(\partial_j \gamma^{kj})R_{ki} {\cal H}
+\Gamma^j_{jk} R^k{}_i {\cal H}
-\Gamma^k_{ji} R^j{}_k {\cal H}
-(\partial_i \gamma^{jk})R_{jk} {\cal H}
\nonumber \\&&
+(\partial_j \gamma^{lj})S^k{}_{li}{\cal M}_k
+\Gamma^j_{jl}S^{kl}{}_{i}{\cal M}_k
-\Gamma^l_{ji} S^{kj}{}_{l}{\cal M}_k
-(\partial_i \gamma^{jl})S^k{}_{jl}{\cal M}_k,
\label{tcma}
\end{eqnarray}
that is,  (\ref{tcha}) and (\ref{tcma})
form a first-order system.
The principal part can be written as
\begin{eqnarray}
\partial_t
\left(
\matrix{{\cal H} \cr {\cal M}_i}\right)
&\simeq&
\left(\matrix{
\beta^l &
-2\alpha\gamma^{jl} \cr
-(1/2)\alpha \delta^l_i
+R^l{}_i
-\delta^l_i R_{km}\gamma^{km}
&
\beta^l \delta^j_i
+S^{jl}{}_i
-\gamma^{mk}\delta^l_iS^j{}_{mk}
}\right)
\partial_l\left(
\matrix{{\cal H} \cr {\cal M}_j}\right). \label{aboveequation}
\end{eqnarray}
The general discussion of the hyperbolicity and characteristic
speed of the system
(\ref{aboveequation}) is hard,
so  hereafter we restrict ourselves to the case
\begin{eqnarray}
R_{ij}=\kappa_1 \alpha \gamma_{ij}
,\quad
S^k{}_{ij}=\kappa_2 \beta^k \gamma_{ij},
\label{res}
\end{eqnarray}
where we recover (\ref{originalADMadjust}) by choosing
$\kappa_1=(\mu-1)/4$ and $\kappa_2=0$.
The eigenvalues of (\ref{aboveequation})  then become
\begin{eqnarray}
\lambda^l&=&
\Big(
\beta^l,\beta^l,(1-\kappa_2)\beta^l \pm
\sqrt{\alpha^2\gamma^{ll}(1+4\kappa_1)+(\kappa_2\beta^l)^2}
\Big)\quad (\mbox{no  sum over } l)
\label{eigen}
\end{eqnarray}
and the hyperbolicity of (\ref{aboveequation})  can be classified as
(i) symmetric hyperbolic when $\kappa_1=3/2$ and $\kappa_2=0$,
(ii) strongly hyperbolic when
$\alpha^2\gamma^{ll}(1+4\kappa_1)+\kappa^2_2(\beta^l)^2 >0$ where
$\kappa_1 \neq -1/4$,
and (iii) weakly hyperbolic when
$\alpha^2\gamma^{ll}(1+4\kappa_1)+\kappa^2_2(\beta^l)^2  \geq 0$.

{}For the case of (\ref{res})
on a Minkowskii background metric,
the linear order terms of the constraint propagation equations become
\begin{eqnarray}
\partial_l\left(
\matrix{{}^{\!(1)\!\!}\hat{\cal H} \cr {}^{\!(1)\!\!}\hat{\cal M}_i}\right)
&=&
\left(
\matrix{0 & -2 i k_j \cr
-(1/2)(1+4\kappa_1)ik_i & 0}
\right)
\left(
\matrix{{}^{\!(1)\!\!}\hat{\cal H} \cr {}^{\!(1)\!\!}\hat{\cal M}_j}\right)
%\partial_t \o {\cal H}&=&
%\beta^j (\partial_j {\cal H})
%-2\alpha\gamma^{ji}(\partial_i {\cal M}_j)
%+2\alpha K(1+2\kappa_1){\cal H}
%+\alpha(\partial_l \gamma_{mn})
%(2\gamma^{ml}\gamma^{nj}-\gamma^{mn}\gamma^{lj}){\cal M}_j
%-4\gamma^{im} (\partial_m\alpha){\cal M}_i
%+4\kappa_2 K \beta^k{\cal M}_k
%\\&\to&
%---------------------------
%\\
%\partial_t \o {\cal M}_i&=&
%-(1/2)\alpha(1+4\kappa_1) (\partial_i {\cal H})
%+\beta^j (\partial_j {\cal M}_i)
%+\alpha K {\cal M}_i
%-(\partial_i\alpha){\cal H}
%+(1-2\kappa_2)(\partial_i\beta^j){\cal M}_j
%-2\kappa_1\alpha \gamma^{mn}(\partial_ i\gamma_{mn}) {\cal H}
%-2\kappa_2(\partial_i\beta^k) {\cal M}_k
%\\&\to&
%-(1/2)(1+4\kappa_1) (\partial_i \o {\cal H}),
\end{eqnarray}
whose Fourier transform gives the eigenvalues
\begin{eqnarray}
\Lambda^l = (0,0,\pm\sqrt{-k^2(1+4\kappa_1)}).
\end{eqnarray}
That is (two 0s, two pure imaginary) for the standard ADM, and
(four 0s) for the original ADM system.
Therefore, according to our conjecture, the standard ADM system is
expected to have
better stability than the original ADM system.

%=====================================================================
\subsection{Detweiler's system}\label{detweilersection}
\subsubsection{Detweiler's system and its constraint amplification}
Detweiler's modification to ADM\cite{detweiler} can be realized
through one
choice  of the multipliers in (\ref{adjADM1}) and (\ref{adjADM2}).
He found that with a particular combination the evolution of the
energy norm of the constraints, ${\cal H}^2+{\cal M}^2$,
can be negative definite
when we apply the maximal slicing condition, $K=0$.
(We will comment more on his criteria
in \S \ref{detweilerdiffsection}.)
His adjustment can be written in our notation in
(\ref{adjADM1}) and (\ref{adjADM2}), as
\begin{eqnarray}
P_{ij}&=&-  L \alpha^3 \gamma_{ij}, \label{Pij_DEThosei}\\
R_{ij}&=&{L } \alpha^3 (K_{ij}- (1 / 3) K \gamma_{ij}), \label{Det1}
\\
S^k{}_{ij}&=&{L } \alpha^2 [
    {3}  (\partial_{(i} \alpha) \delta_{j)}^{k}
%+{3\over 2}  (\partial_j \alpha) \delta_i^{k}
- (\partial_l \alpha) \gamma_{ij} \gamma^{kl}],
\\
s^{kl}{}_{ij}&=&{L } \alpha^3 [
  \delta^k_{(i}\delta^l_{j)}-(1/3) \gamma_{ij}\gamma^{kl}],
  \label{Det3}
\end{eqnarray}
everything else zero, where $L$ is a constant.
Detweiler's adjustment, (\ref{Det1})-(\ref{Det3}),
does not put constraint propagation equation
to first order form, so we can not discuss
hyperbolicity  or the characteristic speed of the constraints.

%\subsubsection{Constraint amplification on Minkowskii background}
{}For the Minkowskii background spacetime,
the adjusted constraint propagation
equations with above choice of multiplier become
\begin{eqnarray}
\partial_l\left(
\matrix{{}^{\!(1)\!\!}\hat{\cal H} \cr {}^{\!(1)\!\!}\hat{\cal M}_i}\right)
&=&
\left(
\matrix{
-2L k^2 & -2 ik_j \cr
-(1/2)ik_i &
-(L/2)k^2 \delta^j_i
-(L/6)k_ik_j}
\right)
%\partial_l
\left(
\matrix{{}^{\!(1)\!\!}\hat{\cal H} \cr {}^{\!(1)\!\!}\hat{\cal M}_j}\right)
%\partial_t \o \gamma_{ij}&=&
%-2L \o {\cal H} \delta_{ij},
%\\
%\mbox{adjust term of }
%\partial_t \o K_{ij}&=&
%+L\partial_{(i} \o {\cal M}_{j)}
%-(L/3)\delta_{ij}(\partial_k \o {\cal M}_k),
%\\
%\partial_t \o {\cal H}&=&
%-2(\partial_j \o {\cal M}_j)
%+4L(\partial_j\partial_j \o {\cal H}),
%\\
%\partial_t \o {\cal M}_i&=&
%-(1/2)(\partial_i \o {\cal H})
%+(L/2)(\partial_k\partial_k \o {\cal M}_i)
%+(L/6)(\partial_i\partial_k \o {\cal M}_k).
\end{eqnarray}
The eigenvalues of the  Fourier transform are
\begin{equation}
\Lambda^l=
\Lambda^l=
(-(L/2)k^2,-(L/2)k^2,
-(4L/3)k^2 \pm \sqrt{k^2(-1+(4/9)L^2 k^2)}).
\end{equation}
\if0
Simplify[Eigenvalues[{
{-4 ell (k1^2+k2^2+k3^2), -2 I k1,-2 I k2,-2 I k3},
{-(1/2) I k1,-(ell/2)(k1^2+k2^2+k3^2)-(ell/6) k1 k1,
  -(ell/6) k1 k2,-(ell/6) k1 k3},
{-(1/2) I k2,-(ell/6) k2 k1,
  -(ell/2)(k1^2+k2^2+k3^2)-(ell/6) k2 k2,-(ell/6) k2 k3},
{-(1/2) I k3,-(ell/6) k3 k1,-(ell/6) k3 k2,
  -(ell/2)(k1^2+k2^2+k3^2)-(ell/6) k3 k3} }]]
\fi
This indicates negative real eigenvalues
if we choose small positive $L$.

We confirmed numerically, using perturbation on Minkowskii,
that Detweiler's system presents better
accuracy than the standard ADM, but only for small positive $L$.
See the Appendix.

%%%%%%%%%%%%%%%%%%
\subsubsection{Differences with Detweiler's requirement}
\label{detweilerdiffsection}
We comment here on the differences between Detweiler's criteria for stable
evolution and ours.

Detweiler calculated the L2 norm of the constraints, $C_\rho$,
over the 3-hypersurface
and imposed the negative definiteness of its evolution,
\begin{equation}
\mbox{Detweiler's criteria}
   \; \Leftrightarrow \;  \partial_t \int  C_\rho C^\rho \ dV <0,
\:  \ \forall \mbox{ non zero } C_\rho.
\end{equation}
where $C_\rho C^\rho =:G^{\rho\sigma} C_\rho C_\sigma$,  and
$G_{\rho\sigma}= diag [1, \gamma_{ij}]$ for the pair of
$C_\rho=({\cal H}, {\cal M}_i)$.

Assuming the constraint propagation to be
$\partial_t \hat C_\rho=A_\rho{}^\sigma \hat C_\sigma$
in the Fourier components,
the time derivative of the L2 norm can be written as
\begin{equation}
\partial_t (\hat{C}_\rho \hat{C}^\rho)
%&=&
%(\partial_t C_a)\bar{C}^a+C_a(\partial_t \bar{C}^a)
%= A^b_a C_b \bar{C}^a
%+C_a (\bar{A}^b_c \bar{C}_b G^{ca}+\bar{C}_b \partial_t \bar{G}^{ba})
=
(A^{\rho\sigma}+\bar{A}^{\sigma\rho}+\partial_t \bar{G}^{\rho\sigma})
\hat{C}_\rho \bar{\hat{C}}_\sigma.
\end{equation}
Together with the fact that the L2 norm is preserved by Fourier
transform,
we can say, for the case of {\it static} background metric,
\begin{equation}
\mbox{Detweiler's criteria}
   \; \Leftrightarrow \;  \mbox{eigenvalues of } (A+A^\dagger)
\mbox{ are all negative} \ \forall k.
\end{equation}

On the other hand,
\begin{equation}
\mbox{Our criteria}
   \; \Leftrightarrow \;  \mbox{eigenvalues of } A
\mbox{ are all negative} \ \forall k.
\end{equation}
%for {\it any} background metric.
Therefore for the case of static background,
Detweiler's criterion is stronger than ours. For example,
the matrix
\begin{equation}
A=
\left(\matrix{
-1&a \cr 0&-1
  }\right)
\mbox{~~where~} a \mbox{~is~constant,}
\end{equation}
for the evolution system $(\hat{C}_1, \hat{C}_2)$
satisfies our criterion
but not Detweiler's
when $|a| \ge \sqrt{2}$.
This matrix however gives asymptotical decay for
$(\hat{C}_1, \hat{C}_2)$.
Therefore we may say that
Detweiler requires the monotonic decay of the constraints,
while we assume only asymptotical decay.

We remark that Detweiler's truncations on higher order terms
in $C$-norm corresponds to our
perturbational analysis;  both are based on the idea that the deviations
from constraint surface
(the errors expressed non-zero constraint value)
are initially small.

\subsection{Another possible adjustment}
\subsubsection{Simplified Detweiler system}
\label{simplifieddetweilersection}
%Due to the discussion in the previous subsection, we can
Similar to Detweiler's (\ref{Pij_DEThosei}),
we next consider only the adjustment
\begin{equation}
P_{ij} = \kappa_0 \alpha \gamma_{ij}, \label{simplifiedDetwadjust}
\end{equation}
all other multipliers zero in (\ref{adjADM1}) and (\ref{adjADM2}).

On the Minkowskii background, the Fourier components of the
constraint propagation equation can be written as
\begin{equation}
\partial_t
\left(\matrix{{}^{\!(1)\!\!} \hat{{\cal H}} \cr {}^{\!(1)\!\!} \hat{{\cal 
M}}_i}\right)
=
\left(\matrix{
2\kappa_0 k^2 & -2ik_j \cr
-(1/2)ik_i &0}\right)
\left(\matrix{{}^{\!(1)\!\!} \hat{{\cal H}} \cr {}^{\!(1)\!\!} \hat{{\cal 
M}}_j}\right),
\end{equation}
and the eigenvalues of the coefficient matrix are
\begin{equation}
\Lambda^l =
(0,0,\kappa_0 k^2 \pm \sqrt{k^2(-1+\kappa_0^2k^2)}).
\end{equation}
That is, the amplification factors become (0, 0, two negative reals)
for the choice of relatively small negative $\kappa_0$.

We also confirmed  that this system works as desired.
We give a  numerical example in the Appendix.

\subsubsection{Adjusting-Hamiltonian-constraints system}
%Due to the discussion in the previous subsection, we can
Our final example is a combination of the one in
\S \ref{section_OriSta} and that above, that is
\begin{eqnarray}
P_{ij}&=&\kappa_0 \alpha \gamma_{ij}, \\
R_{ij}&=&\kappa_1 \alpha \gamma_{ij},
\end{eqnarray}
all other  multipliers zero in (\ref{adjADM1}) and (\ref{adjADM2}).
Similar to the previous one,
the Fourier transformed constraint propagation equation is
\begin{equation}
\partial_t
\left(\matrix{{}^{\!(1)\!\!} \hat{{\cal H}}
\cr {}^{\!(1)\!\!} \hat{{\cal M}}_i}\right)
=
\left(\matrix{
2\kappa_0 k^2 &-2ik_i \cr -(1/2)ik_i-2\kappa_1 ik_i & 0
}\right)
\left(\matrix{{}^{\!(1)\!\!} \hat{{\cal H}}
\cr {}^{\!(1)\!\!}\hat{{\cal M}}_j}\right)
\end{equation}
which gives the eigenvalues
\begin{equation}
\Lambda^l =
(0,0,\kappa_0 k^2 \pm\sqrt{k^2(-1+\kappa_0 k^2-4\kappa_1)}).
\end{equation}
We can expect a similar asymptotical stable evolution
by choosing $\kappa_0$ and  $\kappa_1$, so as to make the
eigenvalues (0, 0, two negative reals).

%\newpage
%=====================================================================
%=====================================================================
%=====================================================================
\section{Conformal-traceless ADM systems} \label{secCTADM}
%=====================================================================
The so-called ``conformally decoupled traceless ADM formulation"
(CT-ADM)
was first developed by the Kyoto group \cite{SN}.
After the re-discovery
that this formulation is more stable than the standard ADM by
Baumgarte and Shapiro \cite{BS}, several groups began to use CT-ADM
formulation for their numerical codes, and reported an advantage
in stability \cite{potsdam0003,LHG}.
Along with this conformal decomposition,
several hyperbolic formulations have
also been proposed \cite{ArBona,FR99,ABMS},
but they have not yet been applied to numerical simulations.

However, it is not yet clear why CT-ADM gives better stability
than ADM.
The Potsdam group \cite{potsdam9908} found that
%the essential point is in
%the process of replacing terms by constraints, and that
the eigenvalues of CT-ADM {\it evolution equations} has fewer
``zero eigenvalues"
than those of ADM, and they conjectured that the instability
can be caused by
``zero eigenvalues" that violate ``gauge mode".
Miller \cite{Miller} applied von Neumann's stability analysis
to the plane wave propagation, and reported that CT-ADM has a
wider range of
parameters that give us stable evolutions.
These studies provide supports of CT-ADM
in some sense, but on the other hand, it is also shown that
an example of an ill-posed solution in CT-ADM (as well in ADM)
\cite{FrittelliGomez}.

Here, we apply our constraint propagation analysis
to this CT-ADM system.

%=====================================================================
\subsection{CT-ADM equations}
%=====================================================================
Since one reported feature of CT-ADM is the use of the momentum
constraint in RHS of the evolution equations \cite{potsdam9908},
we here present the set of CT-ADM
equations carefully for such an replacement of the constraint terms.

The widely used notation\cite{SN,BS} is to use the variables
($\phi,\tilde{\gamma}_{ij}$,$K$,$\tilde{A}_{ij}$,$\tilde{\Gamma}^i$)
instead of the standard ADM ($\gamma_{ij}$,$K_{ij}$), where
\begin{eqnarray}
\tilde{\gamma}_{ij} &=&e^{-4\phi}\gamma_{ij},\label{ctadmval1}
\\
\tilde{A}_{ij} &=&
e^{-4\phi}(K_{ij} - (1/3)\gamma_{ij}K),
\\
\tilde{\Gamma}^i &=& %-\partial_j\tilde{\gamma}^{ij}=
\tilde{\Gamma}^i_{jk}\tilde{\gamma}^{jk}, \label{ctadmval3}
\end{eqnarray}
and we impose ${\rm det}\tilde{\gamma}_{ij}=1$ during the evolutions.
The set of evolution equations become
\begin{eqnarray}
(\partial_t - {\cal L}_\beta) \phi&=&
(-1/6)\alpha K,
\\
(\partial_t - {\cal L}_\beta) \tilde{\gamma}_{ij}&=&
-2\alpha \tilde{A}_{ij},
\\
(\partial_t - {\cal L}_\beta) K
&=&
\alpha (1-\kappa_1)R^{(3)}
+\alpha (1-\kappa_1)K^2
+\alpha\kappa_1 \tilde{A}_{ij}\tilde{A}^{ij}
+(1/3)\alpha\kappa_1 K^2
-\gamma^{ij}(\nabla_i\nabla_j \alpha),
\\
(\partial_t - {\cal L}_\beta) \tilde{A}_{ij}
&=&
-e^{-4\phi} (\nabla_i\nabla_j \alpha)^{TF}
+e^{-4\phi}\alpha R^{(3)}_{ij}
-e^{-4\phi}\alpha (1/3) \gamma_{ij}(1- \kappa_3) R^{(3)}
+\alpha (K \tilde{A}_{ij}- 2 \tilde{A}_{ik} \tilde{A}^k{}_j)
\nonumber \\&&
+e^{-4\phi}\alpha (1/3) \gamma_{ij} \kappa_3
[-\tilde{A}_{kl}\tilde{A}^{kl}+(2/3)K^2],
\\
\partial_t \tilde{\Gamma}^i
&=&
-2(\partial_j \alpha) \tilde{A}^{ij}
-(4/3) \kappa_2\alpha (\partial_j K)\tilde{\gamma}^{ij}
+12\kappa_2\alpha \tilde{A}^{ji}(\partial_j \phi)
-2\alpha \tilde{A}_k{}^j(\partial_j\tilde{\gamma}^{ik})
-2\kappa_2\alpha \tilde{\Gamma}^k{}_{lj}\tilde{A}^j{}_{k}\tilde{\gamma}^{il}
\nonumber \\&&
-2(1-\kappa_2)\alpha 
(\partial_j\tilde{A}_{kl})\tilde{\gamma}^{ik}\tilde{\gamma}^{jl}
+2\alpha(1-\kappa_2) \tilde{A}^i{}_{j}\tilde{\Gamma}^j
\nonumber\\&&
-\partial_j
   \left( \beta^k\partial_k \tilde{\gamma}^{ij}
-\tilde{\gamma}^{kj}(\partial_k\beta^{i})
-\tilde{\gamma}^{ki}(\partial_k\beta^{j})
+(2/3)\tilde{\gamma}^{ij}(\partial_k\beta^k)
           \right)
\end{eqnarray}
where ${\cal L}_\beta$ is the Lie derivative along the shift
vector $\beta^i$, and $R^{(3)}$ is the 3-metric scalar curvature.
Here we introduced
parameters $\kappa$
which show  where we replace the terms with
constraints. For example
$(\kappa_1,\kappa_2, \kappa_3)=(0,0,0)$ is
the case of no replacement
(the standard ADM equations expressed using
(\ref{ctadmval1})-(\ref{ctadmval3})), while Baumgarte-Shapiro \cite{BS} uses
$(\kappa_1, \kappa_2, \kappa_3)=(1,1,0)$.

The constraint equations in CT-ADM system can be expressed as
\begin{eqnarray}
{\cal H}&=&
e^{-4\phi}\tilde R^{(3)}
-8e^{-4\phi}\tilde{\gamma}^{ij}(\partial_i\partial_j\phi)
-8e^{-4\phi}\tilde{\gamma}^{ij}(\partial_i\phi)(\partial_j\phi)
+8e^{-4\phi}(\partial_i\phi)\tilde{\Gamma}^i
+(2/3)K^2
-\tilde{A}_{ij}\tilde{A}^{ij},  \label{CTADM-constraint1}
%-2\Lambda
\\
{\cal M}_i&=&
(\partial_j\tilde{A}_{ki})\tilde{\gamma}^{kj}
-(2/3)(\partial_i K)
-\tilde{A}_{ji}\tilde{\Gamma}^j
+6(\partial_j\phi)\tilde{A}^j{}_i
-\tilde{\Gamma}^k_{ji}\tilde{A}^j{}_{k},  \label{CTADM-constraint2}
\\
{\cal G}^i&=&\tilde{\Gamma}^i+\partial_j\tilde{\gamma}^{ji}.
%{\cal G}^i&=&\tilde{\Gamma}^i-(\partial_j
%\tilde{\gamma}_{mn})\tilde{\gamma}^{mi}\tilde{\gamma}^{nj}.
\label{CTADM-constraint3}
\end{eqnarray}
Here ${\cal H}, {\cal M}$ are the Hamiltonian and momentum constraints
and the third one,
${\cal G}$, is a consistency relation due to the algebraic
definition of (\ref{ctadmval3}).
%[This treatment is different from the one in \cite{FR99}..]

%=====================================================================
\subsection{Constraint propagation equations of CT-ADM }
Similar to the ADM cases, we here show the propagation equations for
(\ref{CTADM-constraint1})-(\ref{CTADM-constraint3}).
The expressions are given using (\ref{d11H}) and (\ref{d11M}), but
we have to be careful to keep using the new variable, $\Gamma_i$,
  wherever it appears.
{}Following \cite{BS}, we express $\tilde R^{(3)}_{ij}$ as
\begin{eqnarray}
\tilde R^{(3)}_{ij}
&=&
-(1/2)\tilde{\gamma}^{lm}(\partial_l\partial_m\tilde{\gamma}_{ij})
+(1/2)\tilde{\gamma}_{ki}\partial_j\tilde{\Gamma}^k
+(1/2)\tilde{\gamma}_{kj}\partial_i\tilde{\Gamma}^k
+(1/2)\tilde{\Gamma}^k \tilde{\Gamma}_{(ij)k}
\nonumber \\&&
+\tilde{\gamma}^{lm}\tilde{\Gamma}^k_{li}\tilde{\Gamma}_{jkm}
+\tilde{\gamma}^{lm}\tilde{\Gamma}^k_{lj}\tilde{\Gamma}_{ikm}
+\tilde{\gamma}^{lm} \tilde{\Gamma}^k_{im}\tilde{\Gamma}_{klj}.
\end{eqnarray}
The constraint propagation equations, then, are obtained by
straightforward calculations as
\begin{eqnarray}
\partial_t {\cal H}&=&
\beta^j (\partial_j {\cal H})
-2\alpha e^{-4\phi}\tilde{\gamma}^{ij}(\partial_i {\cal M}_j)
+2\alpha K{\cal H}
-2\alpha e^{-4\phi}(\partial_i\tilde{\gamma}^{ij}){\cal M}_j
-4\alpha  e^{-4\phi} (\partial_i\phi) \tilde{\gamma}^{ij}{\cal M}_j
-4e^{-4\phi}\tilde{\gamma}^{ij}(\partial_j\alpha){\cal M}_i
\nonumber \\&&
+2\kappa_2 e^{-4\phi}(\partial_i\alpha) \tilde{\gamma}^{ij} {\cal M}_j
+2\kappa_2 e^{-4\phi}\alpha (\partial_i\tilde{\gamma}^{ij}) {\cal M}_j
+2\kappa_2 e^{-4\phi}\alpha \tilde{\gamma}^{ij} (\partial_i{\cal M}_j)
\nonumber \\&&
+16\kappa_2 \alpha e^{-4\phi}(\partial_i\phi)\tilde{\gamma}^{ij} {\cal M}_j
-(4/3)\kappa_1 \alpha K {\cal H},
\\
\partial_t {\cal M}_i&=&
-(1/2)\alpha (\partial_i {\cal H})
+\beta^j (\partial_j {\cal M}_i)
+\alpha K {\cal M}_i
-(\partial_i\alpha){\cal H}
-4\beta^j (\partial_i \phi) {\cal M}_j
+\beta^k \tilde{\gamma}^{jl}(\partial_i\tilde{\gamma}_{lk}){\cal M}_j
+(\partial_i\beta_k)e^{-4\phi}\tilde{\gamma}^{kj}{\cal M}_j
\nonumber \\&&
+(1/3)(2\kappa_1+\kappa_3)(\partial_i \alpha) {\cal H}
+(1/3)(2\kappa_1+\kappa_3)\alpha (\partial_i{\cal H})
-2\kappa_2 \alpha \tilde{A}^j{}_i {\cal M}_j
-(1/3)\kappa_3 \alpha {\cal G}^j \tilde{\gamma}_{ji} {\cal H}
+2\kappa_3 \alpha(\partial_i\phi){\cal H},
\\
\partial_t {\cal G}^i&=&
2\tilde{A}^i{}_j{\cal G}^j
+2\kappa_2 \alpha \tilde{\gamma}^{ij} {\cal M}_j.
\end{eqnarray}
These form a first order system,
and the characteristic part can be extracted as
\begin{eqnarray}
\partial_t
\left(\matrix{{\cal H} \cr {\cal M}_i \cr {\cal G}^i}\right)
&\cong &
\left(\matrix{
\beta^l &2(-1+\kappa_2)\alpha \gamma^{lj}& 0 \cr
((2/3)\kappa_1+(1/3)\kappa_3-(1/2))\alpha \delta^l_i &
\beta^l \delta^j_i & 0 \cr
0&0&0
}\right)
\partial_l\left(\matrix{{\cal H} \cr {\cal M}_j \cr {\cal G}^j}\right),
\label{CTADMconprocha}
\end{eqnarray}
whose characteristic speeds are
\begin{eqnarray}
\lambda^l &=&
\Big(
0,0,0,\beta^l,\beta^l,
\beta^l \pm \alpha
  \sqrt{\gamma^{ll}(1-\kappa_2)(1-(4/3)\kappa_1-(2/3)\kappa_3)}
\Big)  \quad (\mbox{no  sum over } l).
\end{eqnarray}
By analyzing the reality of the eigenvalues,
the diagonalizability of the characteristic matrix,
and the possibility of the
symmetric  characteristic matrix, we can classify
the hyperbolicity of
the system (\ref{CTADMconprocha}) as
\begin{eqnarray}
\mbox{weakly  hyperbolic}
&\Leftrightarrow&
(1-\kappa_2)(1-(4/3)\kappa_1-(2/3)\kappa_3) \geq 0,
\\
\mbox{strongly  hyperbolic}
&\Leftrightarrow&
(1-\kappa_2)=(1-(4/3)\kappa_1-(2/3)\kappa_3)=0,
\nonumber \\&&
\mbox{ or }
(1-\kappa_2)(1-(4/3)\kappa_1-(2/3)\kappa_3) > 0,
\\
\mbox{symmetric  hyperbolic}
&\Leftrightarrow&
(-1+\kappa_2)=(1-(4/3)\kappa_1-(2/3)\kappa_3).
\end{eqnarray}
That is, for the non-adjusted system,
$(\kappa_1,  \kappa_2, \kappa_3)=(0,0,0)$,
constraint propagation forms a strongly hyperbolic system,
while the Baumgarte-Shapiro form gives
only weakly hyperbolicity.
%[Interestingly, this results shows that we can obtain symmetric
%hyperbolic constraint propagation system without
% additional variables
%like \cite{FR99}. ..........]
(We note  that the first-order version of CT-ADM
by Frittelli-Reula \cite{FR99}
has also well-posed constraint propagation equations. )

\subsection{Amplification factors on Minkowskii background}

{}For a Minkowskii background, the constraint propagation
equations at the linear order become
\begin{eqnarray}
\partial_t\left(\matrix{{}^{\!(1)\!\!}\hat{\cal H}
\cr {}^{\!(1)\!\!}\hat{\cal M}_i \cr {}^{\!(1)\!\!}\hat{\cal G}^i}\right)
=
\left(
\matrix{
0 & 2(\kappa_2-1) ik_j & 0 \cr
((2/3)\kappa_1+(1/3)\kappa_3-(1/2))ik_i & 0 &0 \cr
0 & 2\kappa_2 \delta^{ij} & 0 }
\right)
\left(\matrix{{}^{\!(1)\!\!}\hat{\cal H}
\cr {}^{\!(1)\!\!}\hat{\cal M}_i \cr {}^{\!(1)\!\!}\hat{\cal G}^i}\right)
\end{eqnarray}

The constraint amplification factor becomes
\begin{equation}
\Lambda^l=(0,0,0,0,0,
\pm\sqrt{-k^2(1-\kappa_2)(1-(4/3)\kappa_1-(2/3)\kappa_3)})
\end{equation}
That is, $\Lambda^l$ are either zero, pure imaginary or
$\pm$ real numbers.
{}For the non-adjusted system
they are zero and pure imaginary (that is, the same as
(\ref{Beigen_standardADM})),
while the Baumgarte-Shapiro form gives us
all zero eigenvalues.
Therefore from our point of view, these two are not very different
in their characterization of constraint
propagation.

%=====================================================================
\section{Concluding remarks} \label{Summary}
%=====================================================================

We have reviewed ADM systems from the point of view of adjustment
of the dynamical equations by constraint terms.
We have shown that characteristic speeds and amplification factors of the
constraint propagation change due to their adjustments.
We compared the
equations for the ADM, adjusted ADM, conformal traceless ADM (CT-ADM)
systems, and tried to find the system that is robust for violation
of the constraints,
which we can call an ``asymptotically constrained" system.

We conjectured that if the amplification factors
(eigenvalues of the coefficient
matrix of the Fourier-transformed constraint
propagation equations)
are negative or pure-imaginary,
then the system has better asymptotically constrained features
than a system they are not.
According to our conjecture, the standard ADM system is expected
to have
better stability than the original ADM system
(no growing mode in amplification
factors).
Detweiler's modified ADM system,
which is one particular choice of
adjustment,
definitely has good properties in that there
are no growing modes in amplification factors.
We also showed that this can be obtained
by a simpler choice of adjustment multipliers.
%(but turned out that
%this feature is obtained only for the simplified
%range of parameters).

We also studied the CT-ADM system which is popular
with numerical relativists nowadays.
However, from our point of view, we do not see any particular
advantages for CT-ADM system
over the standard ADM system.

The reader might ask why we can break the time-reversal invariant feature of
the evolution equations by a particular choice of adjusting multipliers
against the fact that the ``Einstein equations" are time-reversal invariant.
This question can be answered by the following.
If we take a time-reversal transformation
($\partial_t \, \rightarrow \,
-\partial_t$), the Hamiltonian constraint  and the evolution equations of 
$K_{ij}$
keep their signatures, while the momentum constraints and the evolution
equations of $\gamma_{ij}$ change their signatures.
Therefore if we adjust $\gamma_{ij}$-equations using Hamiltonian constraint
and/or $K_{ij}$-equations using momentum constraints (supposing the
multiplier has $+$-parity), then we can break the time-reversal invariant 
feature
of the ``ADM equations".  In fact, the examples we obtained all
obey  this rule.  The CT-ADM formulation keeps its signature against the
adjustments we made, so that we can not find any additional advantage
from this analysis.

Considering the constraint propagation equations
is a kind of substitutional approach for
numerical integrations of the dynamical equations.
However, this might be one of the
main directions for our future research,
as Friedrich and Nagy \cite{FN} impose the zero
speed of the
constraint propagation as the first principle
when they considered the
initial boundary value problem
of the Einstein equations \cite{Tiglio}.

We are now applying our discussion to more general spacetimes,
and trying to find guidelines for choosing appropriate gauge
conditions from the analysis of the constraint propagation equations.
These efforts will be reported elsewhere
\cite{prep}.

%====================================================================
\section*{Acknowledgements}
%====================================================================
HS appreciates helpful comments
by Pablo Laguna, Jorge Pullin, Manuel Tiglio
and the hospitality of the CGPG group.
We also thank communication with Steven Detweiler.
We thank Bernard Kelly for careful reading of the manuscript.
%Numerical computations were performed using machines at CGPG.
This work was supported in part by the NSF grant PHY00-90091,
and the Everly research funds of Penn State.
HS was supported by the Japan Society for the Promotion of Science
as a research fellow abroad.

%====================================================================
%-------------------------------------------------- appendix   ------
%====================================================================

%====================================================================
\appendix
\section{Numerical demonstrations of adjusted-ADM systems}
\label{appA}

We here show two numerical demonstrations of adjusted-ADM systems
that were discussed in \S \ref{detweilersection}
(Detweiler's modified ADM system)
and \S \ref{simplifieddetweilersection} (simplified version).

  Detweiler's adjustment, (\ref{Det1})-(\ref{Det3}),
can be parametrized by a constant $L$, and our prediction
from the amplification factor on Minkowskii background is that
this system will be asymptotically constrained for
small positive $L$.  Fig.\ref{plot1}
is a demonstration of this system.
We evolved Minkowskii spacetime numerically
in a plane-symmetric spacetime,
and added artificial error in the middle of the evolution.
Our numerical integration uses
the Brailovskaya scheme, which was described in detail in
our previous paper \cite{ronbun1}.
The code passes convergence tests and the plots are for
401 gridpoints in the range $x=[0, 10]$,
and we fix the time grid $\Delta t = 0.2 \Delta x$.
The error was
introduced as a pinpoint kick, in the
form of $\Delta g_{yy}=10^{-3}$ at $x=5.0$ and $t=0.25$.
We monitor how the L2 norm of
the constraints $({\cal H}^2+{\cal M}^2_x)$ behaves.
{}From Fig.\ref{plot1}, we see that a small positive $L$
reduces the L2 norm
in time, which is the asymptotically constrained feature
we expected.
The case of slightly larger $L$ will make the system unstable.
This is the same feature we have seen in the numerical
demonstration of the $\lambda$-system or adjusted-Maxwell/Ashtekar
systems\cite{ronbun2},
for that case the upper bound of the multiplier can be explained by
violation of the
Courant-Friedrich-Lewy condition,
while in this system we can not calculate
the exact characteristics since the system is not first-order.

%>>>>>>>>>>>>>>>>>>>>>>>>>>>>>>>>>>>>>>> Fig.\ref{plot1}
%>>>>>>>>>>>>>>>>>>>>>>>>>>>>>>>>>>>>>>> Fig.\ref{plot1}
\if\answ\nofig
\begin{figure}[h]
\fi
%===========================  figures (one column style) ============
\if\answ\onecol
\begin{figure}[h]
\setlength{\unitlength}{1in}
\begin{picture}(3.3,2.85)
\put(1.5,0.25){\epsfxsize=3.0in \epsffile{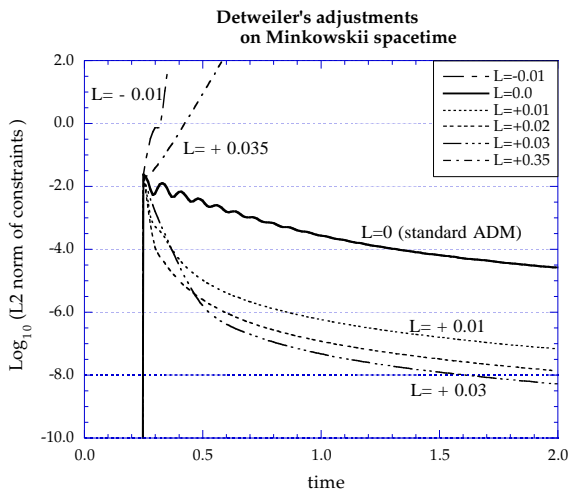} }
\end{picture}
\fi
%===========================  figures (preprint style) ==============
\if\answ\prepri
\begin{figure}[p]
\setlength{\unitlength}{1in}
\begin{picture}(7.0,3.5)
\put(1.0,1.25){\epsfxsize=4.0in \epsfysize=2.38in
\epsffile{plot1.eps}}
\end{picture}
\fi

\caption[degadj]{
Demonstration of the Detweiler's modified ADM system
on Minkowskii background
spacetime (the system of \S \ref{detweilersection}).
The L2 norm of the constraints is plotted in the
function of time. Artificial error was
added at $t=0.25$.
$L$ is the parameter used in (\ref{Det1})-(\ref{Det3}).
We see the evolution is asymptotically constrained for small $L>0$.
}
\label{plot1}
\end{figure}
%<<<<<<<<<<<<<<<<<<<<<<<<<<<<<<<<<<<<<<<     \ref{plot1}
%<<<<<<<<<<<<<<<<<<<<<<<<<<<<<<<<<<<<<<<     \ref{plot1}

Similarly, we plotted in Fig.\ref{plot2} the case of
simplified version
(the system of \S \ref{simplifieddetweilersection}).
We see the desired feature
again by changing the parameter $\kappa_0$ that appear in
(\ref{simplifiedDetwadjust}).

%>>>>>>>>>>>>>>>>>>>>>>>>>>>>>>>>>>>>>>> Fig.\ref{plot2}
%>>>>>>>>>>>>>>>>>>>>>>>>>>>>>>>>>>>>>>> Fig.\ref{plot2}
\if\answ\nofig
\begin{figure}[h]
\fi
%===========================  figures (one column style) ============
\if\answ\onecol
\begin{figure}[h]
\setlength{\unitlength}{1in}
\begin{picture}(3.3,2.85)
\put(1.5,0.25){\epsfxsize=3.0in \epsffile{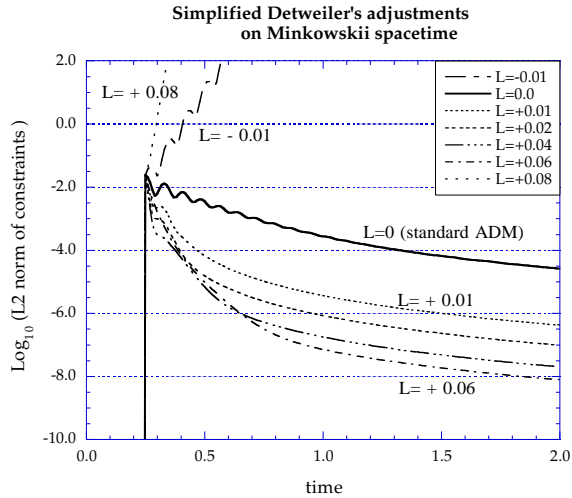} }
\end{picture}
\fi
%===========================  figures (preprint style) =============
\if\answ\prepri
\begin{figure}[p]
\setlength{\unitlength}{1in}
\begin{picture}(7.0,3.5)
\put(1.0,1.25){\epsfxsize=4.0in \epsfysize=2.38in
\epsffile{plot2.eps} }
\end{picture}
\fi

\caption[degadj]{
Demonstration of the simplified Detweiler's modified ADM system
on Minkowskii background
spacetime (the system of \S \ref{simplifieddetweilersection}).
%The L2 norm of the constraints is plotted in the
%function of time.
{}For comparison with Fig.\ref{plot1}, we set
$L=-\kappa_0$, where $\kappa_0$ is the parameter used
in (\ref{simplifiedDetwadjust}).
We see the evolution is asymptotically constrained for small $L>0$.
}
\label{plot2}
\end{figure}
%<<<<<<<<<<<<<<<<<<<<<<<<<<<<<<<<<<<<<<<     \ref{plot2}
%<<<<<<<<<<<<<<<<<<<<<<<<<<<<<<<<<<<<<<<     \ref{plot2}

\newpage
%====================================================================
%====================================================================

%-------------------------------------------------- references ------
%====================================================================

\end{document}